\documentclass[showpacs,aps,twocolumn]{revtex4}

\usepackage{bm}
\usepackage{amsmath}
\usepackage{graphicx}
\usepackage{subfigure}
\usepackage[usenames,dvipsnames]{color}
\definecolor{darkblue}{RGB}{0,0,196}
\usepackage[colorlinks=true,linkcolor=darkblue,citecolor=darkblue,urlcolor=darkblue]{hyperref}
\usepackage{array}

\usepackage{setspace}
\usepackage{footmisc}
\usepackage[makeroom]{cancel}
\usepackage{comment}

\usepackage{color,soul}

\def\be{\begin{equation}}
\def\ee{\end{equation}}
\def\ba{\begin{eqnarray}}
\def\ea{\end{eqnarray}}

\usepackage{graphicx}
\usepackage{amsmath,bbm}
\usepackage{amssymb,bm}

\begin{document}
\title{ Thermodynamic and transport properties in $Au+Au$ collisions at RHIC energies from the clustering of color strings}

\author{Pragati Sahoo}
\author{Swatantra~Kumar~Tiwari}
\author{Sudipan De}
\author{Raghunath~Sahoo\footnote{Corresponding author: $Raghunath.Sahoo@cern.ch$}}
\affiliation{Discipline of Physics, School of Basic Sciences, Indian Institute of Technology Indore, Indore- 453552, INDIA}
\author{Rolf P. Scharenberg}
\author{Brijesh K. Srivastava}
\affiliation{Department of Physics and Astronomy, Purdue University, West Lafayette, IN-47907, USA}
\begin{abstract}
\noindent

In this work, we have extracted the initial tempearture from the transverse momentum spectra of charged particles in $Au+Au$ collisions using  STAR data at RHIC energies from $\sqrt{s_{NN}}$ = 7.7 to 200 GeV. The initial energy density ($\varepsilon$), shear viscosity to entropy density ratio ($\eta/s$), trace anomaly ($\Delta$), the squared speed of sound ($C_s^2$), entropy density, and bulk viscosity to entropy density ratio ($\zeta/s$) are obtained and compared  with the lattice QCD calculations for (2+1) flavor. The initial temperatures obtained are compared with various hadronization and chemical freeze-out temperatures. The analysis of the data shows that the deconfinement to confinement transition possibly takes place between $\sqrt{s_{NN}}$ = 11.5 and 19.6 GeV.

\end{abstract}

\date{\today}
\maketitle 
\section{Introduction}
\label{intro}
 One of the main goals of relativistic heavy ion collision is to study the deconfined matter, known as Quark-Gluon Plasma (QGP), which is expected to be formed at very high temperatures. The Relativistic Heavy Ion Collider (RHIC) at Brookhaven National Laboratory is one of the major facilities that can explore the QCD phase diagram. Theoretical models, based on lattice QCD (lQCD), predict that at vanishing baryon chemical potential ($\mu_{B}$), the transition from QGP to hadron gas is a smooth crossover~\cite{crossover}, while several QCD based calculations show that at larger $\mu_{B}$ a first order phase transition may take place. The Beam Energy Scan (BES) program at RHIC is dedicated to locate the QCD critical point. In BES program, RHIC has collided heavy-ion (Au+Au) beams at $\sqrt{s_{NN}}$ = 7.7 - 200 GeV. 

 Theoretically, several signatures of first order phase transition and the critical point have been proposed~\cite{cp1, cp2, cp3}. A non-monotonic variation of conserved quantum number fluctuations as a function of $\sqrt{s_{NN}}$ is found near the possible critical point~ \cite{netproton}. Transport properties of strongly interacting matter, such as shear and bulk viscosities are of particular importance to understand the nature of QCD matter. It is expected that the ratio of shear viscosity ($\eta$) to entropy density ($s$) would exhibit a minimum value near the QCD critical point~\cite{etabys}. Recently, the STAR experiment has reported some interesting features regarding critical point around $\sqrt{s_{NN}}$ = 19.6 GeV~\cite{netproton}. The higher moments of net-proton distribution show significant deviation from Poissonian expectation and Hadron Resonance Gas model prediction at $\sqrt{s_{NN}}$ = 19.6 and 27 GeV~\cite{netproton}. Also, the two particle transverse momentum ($p_{T}$) correlation scaled with average transverse momentum ($\langle p_{T} \rangle$) fluctuation, which is related to the specific heat, $C_{V}$ of the system, significantly decreases below $\sqrt{s_{NN}}$ = 19.6 GeV~\cite{ptfluc}. Therefore, it would be very interesting to study the thermodynamical quantities and transport properties of the QCD matter with special emphasis on RHIC BES energies.

 The article is organised as follows. In section~\ref{string} the dynamics of string interactions and in section~\ref{FM} the formalism and methodology of CSPM are discussed. Section~\ref{TTP} covers a study of 
 the initial temperatures at RHIC energies as obtained using CSPM  and a comparison with various chemical
freeze-out results. In addition, the thermodynamical and transport quantities like energy density, shear viscosity, trace anomaly, speed of sound, entropy density, bulk viscosity of the matter produced in heavy-ion collision at RHIC by using the CSPM are discussed.  Finally in section~\ref{summary}, we present the summary and conclusions.

\section{String Interactions And Dynamics}
\label{string}

From theoretical point of view, in addition to hydrodynamic studies, Color Glass Condensate (CGC), which is derived directly from QCD, gives reasonable description of several experimental observables~\cite{CGC}. An alternative approach to the CGC is the percolation of strings~\cite{Phyreport}. It is a QCD inspired model but is not directly obtained from QCD. In clustering of color sources approach, the color flux tubes or color strings are stretched between the colliding partons in terms of the color field. The strings produce ${\it q\bar q}$ pair in finite space filled with the chromoelectric field similar to the Schwinger mechanism of pair creation in a constant electric field covering all the space \cite{Phyreport}. The number of strings grows with the energy and with the number of nucleons of participating nuclei. Color strings may be viewed as small discs in the transverse space filled with the color field created by colliding partons. With growing energy and size of the colliding nuclei the number of strings grows and start to overlap and interact to form clusters, in the transverse plane very much similar to disks in two dimensional (2D) percolation theory~\cite{Isichenko:1992zz}. 

At a critical string density a macroscopic cluster appears that marks the percolation phase transition which spans the transverse nuclear interaction area. This is termed as Color String Percolation Model (CSPM). The general result, due to the SU(3) random summation of charges, is a reduction in multiplicity and an increase in the string tension hence increase in the average transverse momentum squared, $\langle p_{T}^{2} \rangle$ \cite{Phyreport}.  

However in CSPM the Schwinger barrier penetration mechanism for particle production, the fluctuations in the associated string tension and taking into account the quantum fluctuations of the color field make it possible to define a temperature. Consequently the particle spectrum is ``born" with a thermal distribution. When the initial density of interacting colored strings ($\xi$) exceeds the 2D percolation threshold ($\xi_c$) i. e. $\xi > \xi_c $, a macroscopic cluster appears, which defines the onset of color deconfinement. This happens at $ \xi_c \ge 1.2$ \cite{Isichenko:1992zz,Satz:2000bn}. The critical density of percolation is related to the effective critical temperature and thus percolation may be the way to achieve deconfinement in the heavy-ion collisions \cite{PLB642}. It is observed that, CSPM can be successfully used to describe the initial stages in high energy heavy-ion collisions \cite{Phyreport}. In the present work, the thermodynamical variables and transport coefficients are obtained using CSPM. The results are compared with lQCD predictions~\cite{Bazavov:2009zn}. CSPM has been successfully applied to small systems as well. It has been shown that de-confinement can be achieved in high multiplicity events $\bar{p}p$ collisions at $\sqrt{s}$ = 1.8 TeV in E-735 experiment \cite{Gutay:2015cba}. Thus CSPM is a new paradigm which has been successful in explaining the initial thermalization both in $A+A$ and in high multiplicity $\bar{p}p$ collisions at  $\sqrt{s}$ = 1.8 TeV.

\section{Formulation And Methodology}
\label{FM}
In CSPM, the interactions of strings or in other words the overlapping of strings reduces the hadron multiplicity ($\mu$) and increases the average transverse momentum squared, $\langle p_{T}^2 \rangle$ of these hadrons to conserve the total transverse momentum. The hadron multiplicity and $\langle p_{T}^2 \rangle$ are directly related to the field strength of the color sources and thus to the generating color. For a cluster of $n$ individual strings, we have \cite{epjc71},   

\begin{eqnarray}
n = \frac{\mu}{\mu_0}\frac{\langle p_{T}^2 \rangle}{\langle p_{T}^2 \rangle _1 },
\end{eqnarray}
where $\mu_0$ and $\langle p_{T}^2 \rangle _1$ are the multiplicity and the mean transverse momentum squared of particles produced from a single string \cite{epjc71}, respectively. As the number of strings, $n$ increases the macroscopic cluster suddenly spans the area. In 2D percolation theory, the dimensionless percolation density parameter is given by~\cite{prc65, epjc16}

\begin{eqnarray}
\xi = \frac{N_{S}S_1}{S_n} ,
\end{eqnarray}
where $N_S$ and $S_n$ being the total number of individual strings and interaction area, respectively. $S_1$ is the transverse area of a single string. We evaluate the initial value of $\xi$ by fitting the experimental data of $p_{T}$ spectra in $pp$ collisions at $\sqrt{s}$ = 200 GeV using the following function:
\begin{eqnarray}
\frac{dN_{c}}{dp_{T}^{2}} = \frac{a}{(p_{0}+{p_{T}})^{\alpha}},
\end{eqnarray}
where, {\it a} is the normalisation factor and  $p_{0}$, $\alpha$ are fitting parameters given as, $p_{0}$ = 1.982 and $\alpha$ = 12.877 \cite{Phyreport}. In order to evaluate the interactions of strings in A+A collisions, we use the above parameterisation as follows: 

\begin{eqnarray}
 p_0\rightarrow p_0\left(\frac{\langle  nS_1/S_n\rangle_{AuAu}}{\langle nS_1/S_n\rangle_{pp}}\right)^{1/4}.
 \label{po}
\end{eqnarray}

Here, $S_n$ corresponds to the area occupied by the $n$ overlapping strings. Using thermodynamic limit, {\it i.e.} $n$ and $S_n \rightarrow \infty$ and keeping $\xi$ fixed, we get

\begin{eqnarray}
 \langle \frac{nS_{1}}{S_{n}} \rangle = \frac{1}{F^{2}(\xi)},
 \label{p1}
\end{eqnarray}
where $F(\xi)$ is the color suppression factor which reduces the hadron multiplicity from $n\mu_0$ to the interacting string value, $\mu$ as 

\begin{eqnarray}
\mu = F(\xi)n\mu_0,
\end{eqnarray}
where,
\begin{eqnarray}
 F(\xi) = \sqrt \frac{1-e^{-\xi}}{\xi}.
 \label{p1}
\end{eqnarray}

Using Eq.(3), we obtain for $A+A$ collisions as,
\begin{eqnarray}
\frac{dN_{c}}{dp_{T}^{2}} = \frac{a}{(p_{0} \sqrt {F(\xi)_{pp}/F(\xi)_{AuAu}}+{p_{T}})^{\alpha}}.
\end{eqnarray}

Here, $\langle p_{T}^{2} \rangle_{n}$ = $\langle p_{T}^{2} \rangle_{1}$/F($\xi$). $\mu$ is the multiplicity and $\langle p_{T}^{2} \rangle_{n}$ is the mean transverse momentum squared of the particles produced by a cluster of $n$ strings.
In $pp$ collisions, $\langle nS_{1}/S_{n}\rangle \sim 1$ due to the low string overlap probability. The measured values of $\xi$ for different RHIC energies are tabulated in Table~\ref{table}.

The initial temperature of the percolation cluster, $T(\xi)$ can be represented in terms of F($\xi$) as~\cite{Phyreport}: 
\begin{eqnarray}
T(\xi) = \sqrt \frac{\langle p_{T}^{2} \rangle_{1}}{2F(\xi)}.
\label{temp}
 \end{eqnarray}		

Recently, it has been suggested that fast thermalization in heavy ion collisions can occur through the existence of an event horizon caused by a rapid deceleration of the colliding nuclei \cite{khar2}. The thermalization in this case is due to the Hawking-Unruh effect \cite{hawk,unru}. In CSPM the strong color field inside the large cluster produces deceleration of the primary $q \bar q$ pair which can be seen as a thermal temperature by means of the Hawking-Unruh effect.   

The single string squared-average transverse momentum, $\langle p_{T}^{2} \rangle_{1}$ is calculated using eq.~(\ref{temp}) at critical temperature, $T_{c}$ = 167.7 $\pm$ 2.76 MeV \cite{Becattini:2010sk} and $\xi_c \sim$ 1.2. We get $\sqrt{\langle p_{T}^{2} \rangle_{1}}$ = 207.2$\pm$3.3 MeV~\cite{Phyreport}, which is $\simeq$ 200 MeV, obtained in the previous calculation using percolation model~\cite{PLB642}. The initial temperatures ($T$) for different energies corresponding to their $\xi$ values are also tabulated in Table~\ref{table1}. 

\begin{figure}
\includegraphics[height=20em]{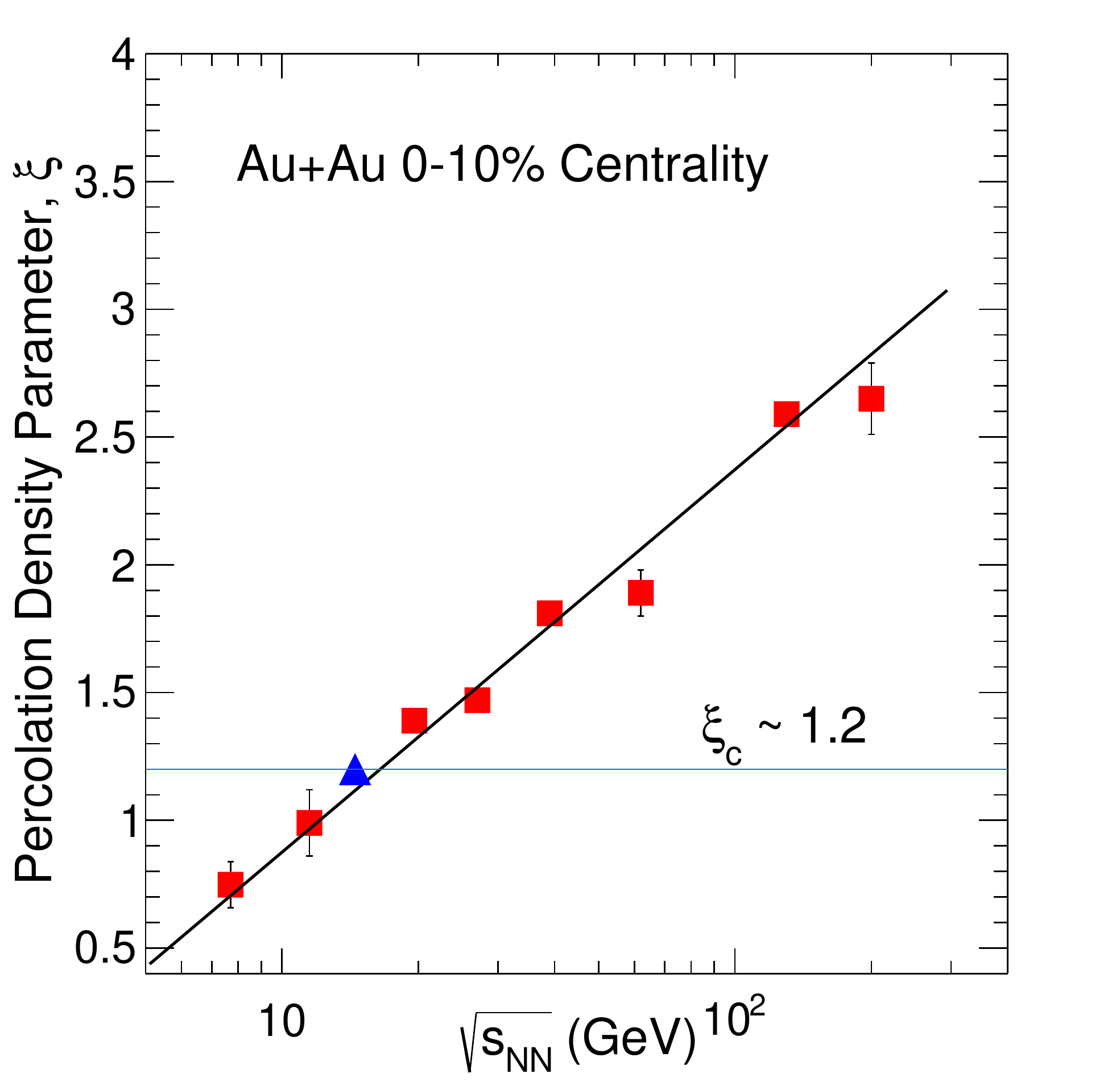}
\caption[]{(Color online) Percolation density parameter, $\xi$ as a function of $\sqrt{s_{NN}}$. The red squares show the values for RHIC energies from $\sqrt{s_{NN}}$ = 7.7 - 200 GeV. The blue triangle is the prediction for 14.5 GeV. The horizontal line at $\xi \sim$ 1.2 is the critical value of $\xi$ \cite{Isichenko:1992zz,Satz:2000bn}.}
\label{xiecm}
\end{figure}

Figure~\ref{xiecm} shows the percolation density parameter $\xi$ as a function of beam energy. It is observed that $\xi$ is a linear function of $\sqrt{s_{NN}}$. The horizontal line in Fig.~\ref{xiecm} at $\xi_{c}$ = 1.2 is the percolation threshold at which the spanning cluster appears, a connected system of color sources and identifies the percolation phase transition \cite{Isichenko:1992zz,Satz:2000bn}. It is observed that this threshold is achieved for $\sqrt{s_{NN}}$ = 19.6 GeV and above.

\begin{table*}[tp]
 \centering
  \caption{The thermodynamical observables and transport coefficients estimated in CSPM for ($0 - 10$)\% central Au+Au collisions at various RHIC energies.}
  \label{table}

\begin{tabular} {| m{1.5cm} | m{1.5cm} | m{1.5cm} | m{1.5cm} | m{1.5cm} | m{1.5cm} | m{1.5cm} | m{1.5cm} | m{1.5cm} | m{1.5cm} | m{1.5cm} | m{1.5cm} |}
 \hline
$\sqrt{s_{NN}}$ (GeV) & $\xi$  & $F(\xi) $   & $\varepsilon/T^{4}$    & $c_s^2$ &        $s/T^{3}$         & $\eta/s$           & $\zeta/s$        & $\Delta $ & $T$ (MeV)     \\
\hline
$7.7$          & 0.75 $\pm$ 0.09    & 0.84 $\pm$ 0.03     & 6.90 $\pm$ 1.16   & 0.09 $\pm$ 0.04        & 7.56 $\pm$ 1.30             & 0.31 $\pm$ 0.03     & 0.26 $\pm$ 0.09        & 3.25 $\pm$ 0.30  & 159.96 $\pm$ 3.82  \\
\hline
$11.5$        & 0.99  $\pm$ 0.13   & 0.80 $\pm$ 0.03      & 8.24  $\pm$ 1.42   & 0.12 $\pm$ 0.04        & 9.23 $\pm$ 1.63          & 0.26 $\pm$ 0.02     & 0.18  $\pm$ 0.07   &3.78 $\pm$ 0.31  & 164.16 $\pm$ 4.05 \\
\hline
$19.6$         &  1.39 $\pm$ 0.05    & 0.740 $\pm$ 0.006     & 9.85  $\pm$ 0.85    & 0.160 $\pm$ 0.009        & 11.41   $\pm$ 0.99           & 0.230 $\pm$ 0.005       & 0.106 $\pm$ 0.011 & 4.34$\pm$ 0.10   & 170.83 $\pm$ 2.82  \\
\hline
$27$           &  1.47 $\pm$ 0.03   & 0.724 $\pm$ 0.003     & 10.10   $\pm$ 0.80    & 0.165   $\pm$ 0.005       & 11.77  $\pm$ 0.94            & 0.227 $\pm$ 0.004     & 0.097 $\pm$ 0.005  & 4.41 $\pm$ 0.08   & 172.15 $\pm$ 2.76      \\
\hline
$39$           &  1.81$\pm$ 0.05    & 0.679 $\pm$ 0.006     & 10.95   $\pm$ 0.92    & 0.190  $\pm$ 0.006       & 13.04  $\pm$ 1.10           & 0.215 $\pm$ 0.004     & 0.066 $\pm$ 0.006  & 4.64 $\pm$ 0.09  & 177.72  $\pm$ 2.95      \\
\hline
$62.4$         &  1.89 $\pm$ 0.09   & 0.67 $\pm$ 0.01     & 11.14  $\pm$ 1.07      & 0.19  $\pm$ 0.01       & 13.32  $\pm$ 1.28          & 0.213 $\pm$ 0.005    & 0.06 $\pm$ 0.01   & 4.68 $\pm$ 0.11  & 178.89 $\pm$ 3.14    \\
\hline
$130$          & 2.59 $\pm$ 0.04    & 0.597  $\pm$ 0.004    & 12.10  $\pm$ 0.97    & 0.234  $\pm$ 0.003       & 14.94  $\pm$ 1.19          & 0.207 $\pm$ 0.003     & 0.030 $\pm$ 0.001   & 4.81 $\pm$ 0.08  & 189.59  $\pm$ 3.07    \\
\hline
$200$          &  2.65 $\pm$ 0.14   & 0.59 $\pm$ 0.01      & 12.19  $\pm$ 1.23   & 0.24     $\pm$ 0.01    & 15.08    $\pm$ 1.53      & 0.207 $\pm$  0.004 & 0.028 $\pm$ 0.006    & 4.82 $\pm$ 0.10  & 190.32   $\pm$ 3.50    \\

\hline
\end{tabular}
\label{table1}
\end{table*}

\section{Thermodynamic and transport properties}
\label{TTP}

In this work, we attempt to determine the thermodynamic and transport properties of the strongly interacting matter produced in the central Au+Au collisions at various RHIC energies ranging from 7.7 to 200 GeV using CSPM. We study the thermodynamical properties such as, energy density, speed of sound, entropy density and transport properties like shear and bulk viscosities as discussed below.

\subsection{Chemical Freeze-out and Initial Temperature}
\label{CF}
\begin{figure}
\includegraphics[height=20em]{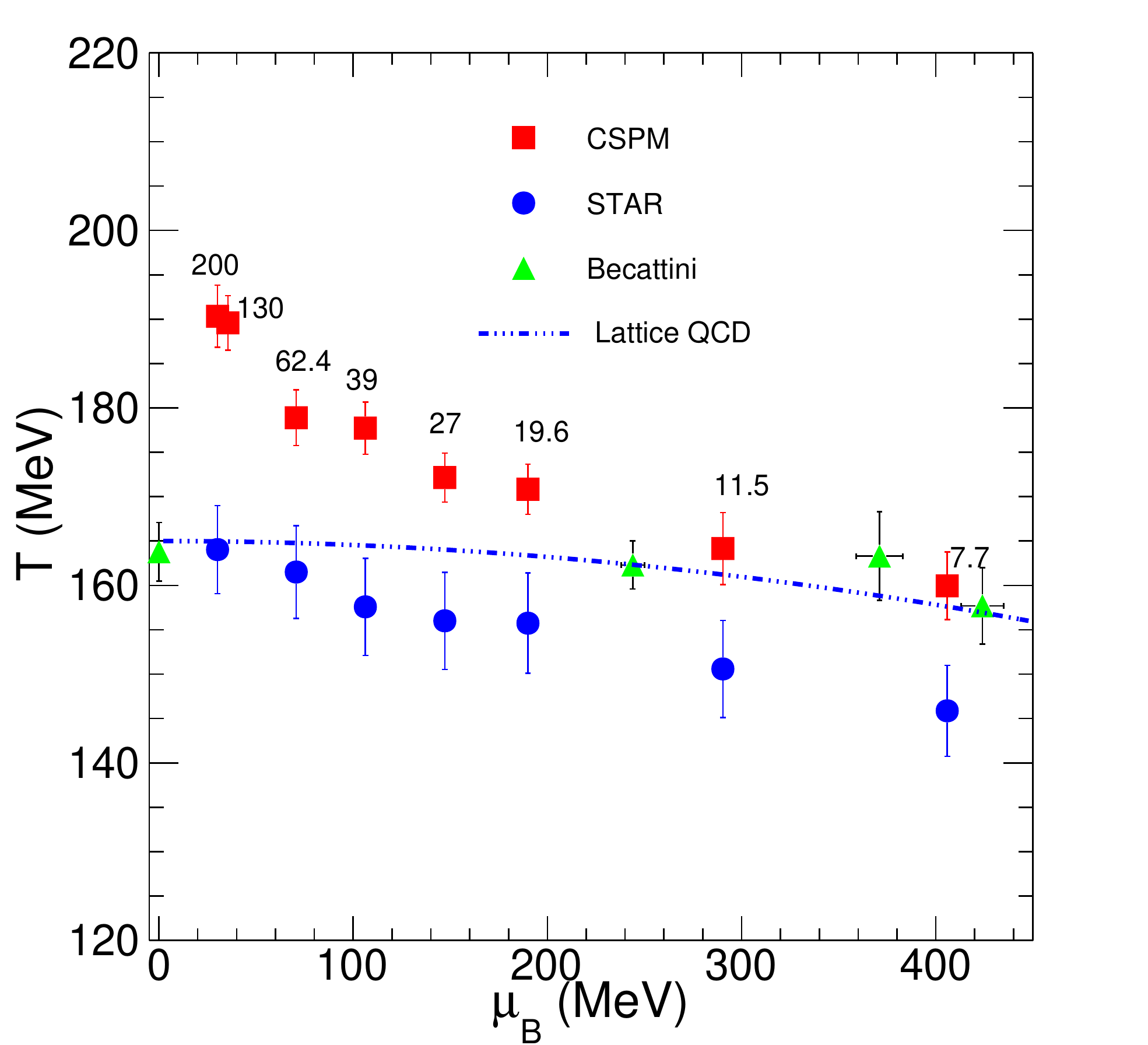}
\caption[]{(Color online) The temperature and baryon chemical potential ($\mu_B$) estimated in different calculations at various center-of-mass energies. The red squares are the initial temperatures at RHIC energies estimated in CSPM. The blue dash-dotted line is the prediction from lQCD~\cite{Bazavov:2017dus}. The green triangles are for hadronization temperature and baryon chemical potential as obtained in the statistical model~\cite{Becattini:2016xct}. The blue circles are (T, $\mu_B$) at freeze-out estimated by STAR experiment~ \cite{Das:2014qca}.}
\label{Tmu}
\end{figure}

The predictions of chemical freeze-out temperature ($T_{ch}$), and baryon chemical potential ($\mu_B$) obtained at various energies of RHIC experiment are shown in Figure \ref{Tmu}. The blue dash-dotted line represents the lQCD calculations of the chiral curvature in terms of T and $\mu_B$~\cite{Bazavov:2017dus}. The coordinates of hadronization points estimated by using a transport model fit to the experimental data at SPS energies are shown by green triangles \cite{Becattini:2016xct}. The blue circles are the experimentally measured values of ($T_{ch}$, $\mu_B$) by the STAR experiment \cite{Das:2014qca}, which uses a statistical thermal model fit to the experimental particle ratios. The initial temperatures obtained in CSPM at RHIC energies are presented by red squares and it is found that as we go from lower energies to higher energies, the differences between initial temperatures and chemical freeze-out temperatures increase. The initial energy density created in heavy-ion collisions govern the subsequent hadronization and system evolution to the final state, characterized by the system freeze-out.  The large difference in the initial temperature at higher energies, as obtained in the framework of CSPM and the chemical freeze-out temperatures obtained in the framework of statistical hadron gas models using experimental particle ratios are simply because of the higher initial energy densities (temperatures) at lower baryochemical potentials. The values of $T_{ch}$ obtained by STAR experiment lie below the lQCD results except at vanishing baryon chemical potential. 

\subsection{Initial Energy Density}
CSPM assumes that the initial temperature of the fluid in local thermal equilibrium is determined at the string level. So, CSPM along with boost invariant Bjorken hydrodynamics~\cite{Bjorken:1982qr} is used to calculate energy density, pressure, entropy etc. The expression for initial energy density ($\varepsilon$) is as follows :
 
\begin{eqnarray}
 \varepsilon = \frac{3}{2}\frac{\frac{dN_{c}}{dy}\langle m_{T} \rangle}{S_{N}\tau_{pro}},
 \label{p1}
\end{eqnarray}

where $S_{N}$ is the nuclear overlap area estimated by using Glauber model \cite{star1} and $\tau_{pro}$ is the production time for a boson (gluon). Here, $m_T$ = $\sqrt{m^2+p_T^2}$ is the transverse mass. For evaluating $\varepsilon$, we use the charged pion multiplicity $dN_c/dy$ at midrapidity and $S_n$ values from STAR for $0\%-10\%$ central Au+Au collisions at $\sqrt{s_{NN}}$ = 7.7 - 200 GeV~\cite{star1, star2}. The dynamics of massless particle production has been studied in two-dimensional Quantum Electrodynamics (QED$_2$). QED$_2$ can be scaled from electrodynamics to quantum chromodynamics using the ratio of the coupling constants~\cite{qe2}. The production time ($\tau_{pro}$) for the boson (gluon) is~\cite{taopro} 

\begin{eqnarray}
 \tau_{pro} = \frac{2.405\hbar}{\langle m_{T} \rangle}.
 \label{tau}
\end{eqnarray}

\begin{figure}
\includegraphics[height=20em]{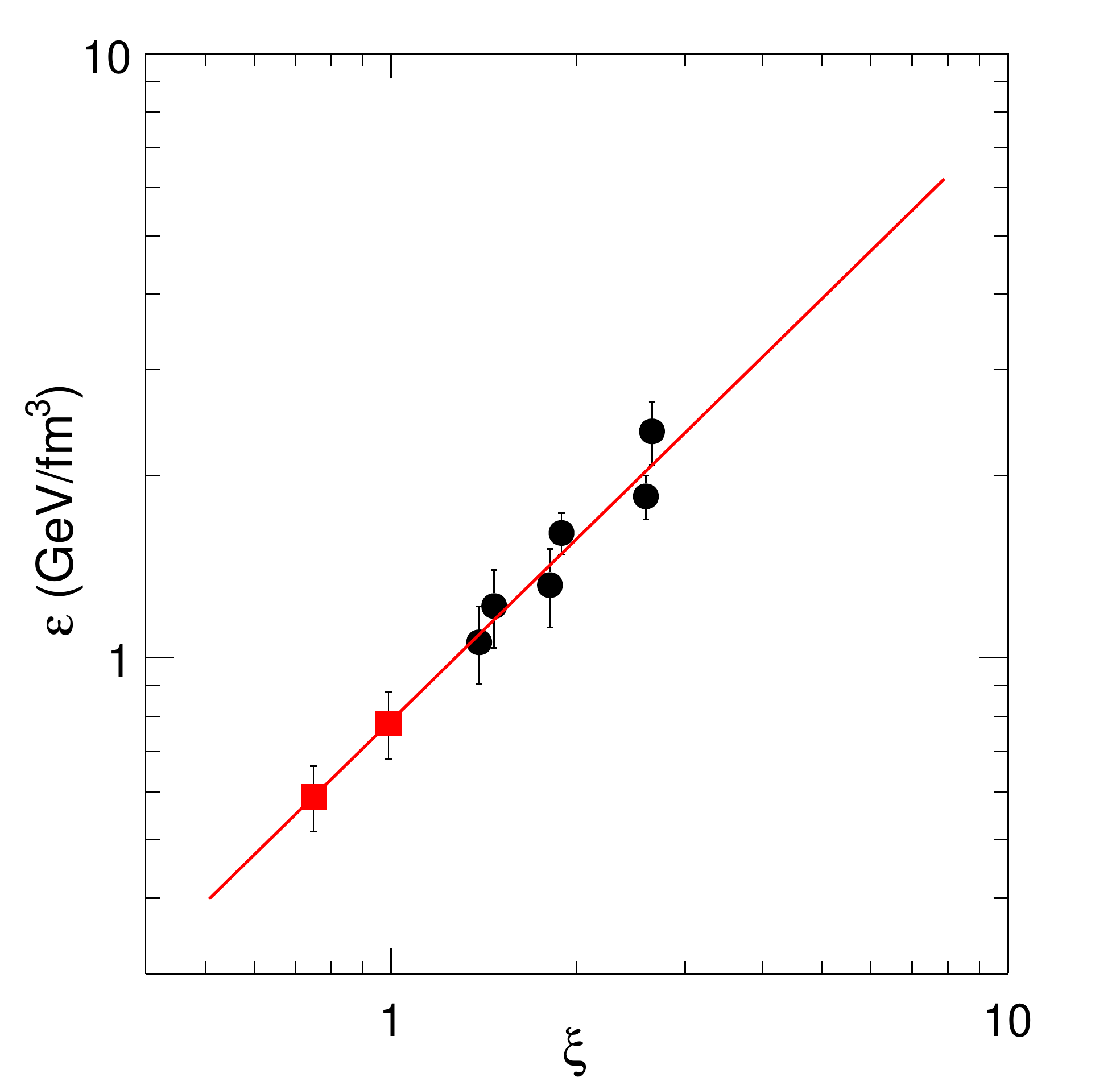}
\caption[]{(Color online) Initial energy Density  $\varepsilon$ as a function of the percolation density parameter, $\xi$. The black circles show the values for RHIC energies from $\sqrt{s_{NN}}$ = 19.6 - 200 GeV. The red squares are the interpolated $\xi$ values for 7.7 and 11.5 GeV. }
\label{xi}
\end{figure}

Figure~\ref{xi} shows the variation of the energy density with $\xi$. Solid circles are the results obtained using CSPM at RHIC energies starting from $\sqrt{s_{NN}}$ = 19.6 - 200 GeV. Line represents fitting of CSPM results. Here, we do not consider the results at $\sqrt{s_{NN}}$ = 7.7, and 11.5 GeV as they behave differently in comparison to the other RHIC energies due to the excess presence of baryons. It is found that $\varepsilon$ is proportional to $\xi$. The parameterisation of the CSPM results gives,  $\varepsilon$ = 0.786 $\xi$ (GeV/$fm^3$). We extrapolate this relationship to estimate $\xi$ values for $\sqrt{s_{NN}}$ = 7.7, and 11.5 GeV. We use this relationship to calculate various thermodynamical and transport properties.

\begin{figure}
\includegraphics[height=20em]{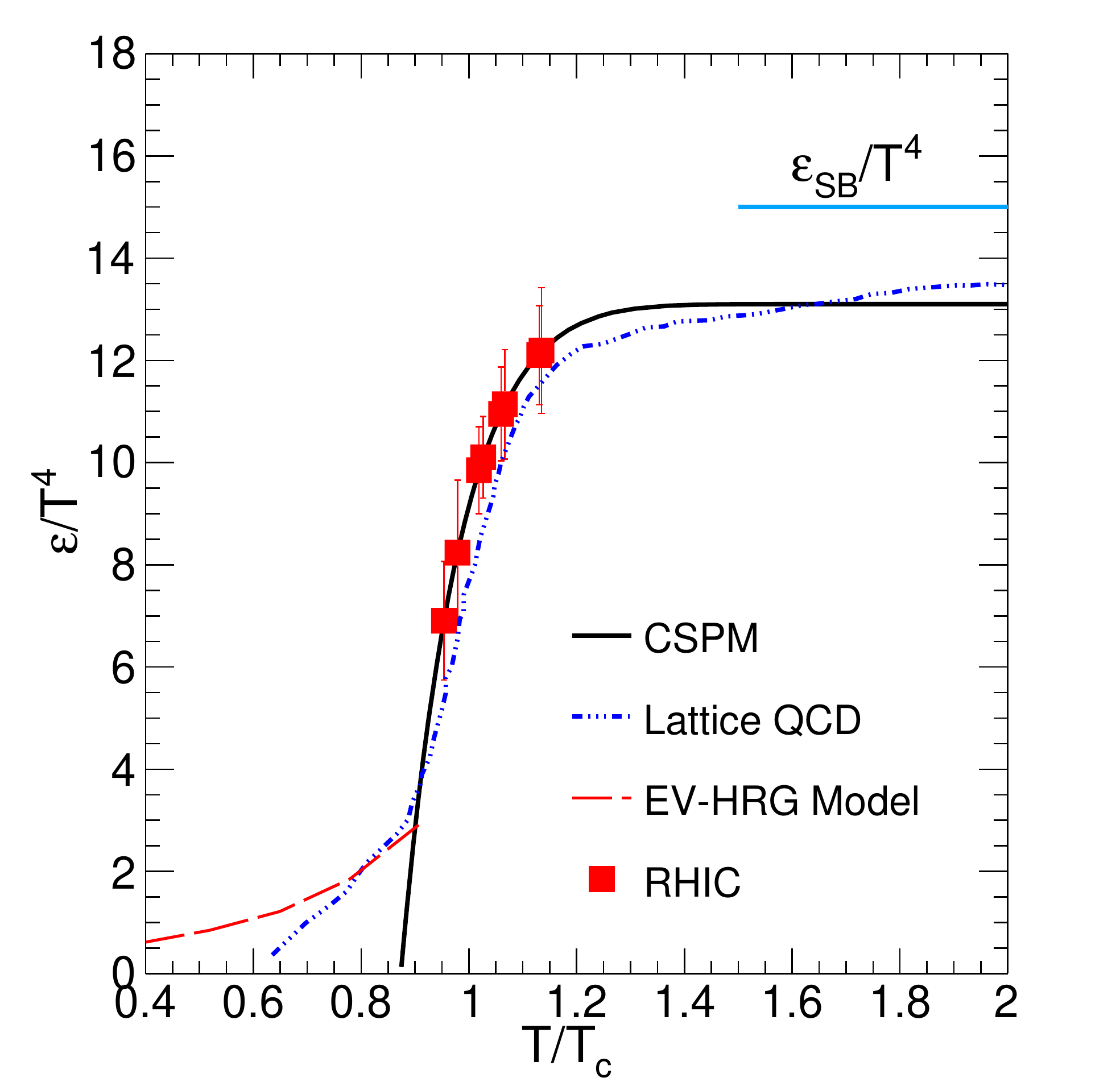}
\caption[] {(Color online) Scaled energy density ($\varepsilon/T^{4}$) as a function of $T/T_{c}$. The black solid line is CSPM results and the blue dash-dotted line corresponds to the lattice QCD results~\cite{Bazavov:2009zn}. The red dashed line shows the result from EV-HRG model~\cite{Tiwari:2011km}.}
\label{Edensity}
\end{figure}

Figure~\ref{Edensity} presents the variation of  $\epsilon/T^4$, which is proportional to degrees of freedom of the system with scaled temperature (T/T$_{c}$), where $T_c$ is the critical temperature. The square symbols are the results calculated from CSPM at various RHIC energies starting from 7.7 - 200 GeV. The lattice QCD results from HotQCD collaboration~\cite{Bazavov:2009zn} are also shown by the blue dash-dotted line. The solid line is the CSPM results calculated using our parameterisation. The red dashed line shows the result for hadron gas calculated using Excluded-Volume Hadron Resonance Gas (EV-HRG) model~\cite{Tiwari:2011km} at $\mu_B$ = 0, which matches with the lQCD results at lower temperature. We find that $\varepsilon/T^4$ varies rapidly around T/T$_{c} \sim$  1, which suggests that the number of degrees of freedom change rapidly at this temperature and there is a cross-over phase transition from hadron gas to quark-gluon plasma. After T/T$_{c}\sim$ 1.2, we observe that $\varepsilon/T^4$ saturates with temperature. It is observed that CSPM results are in good agreement with the lattice QCD results~\cite{Bazavov:2009zn} and always lie below to the value of the ideal gas Stefan-Boltzmann limit.

\subsection{Shear Viscosity}

The observation of the large elliptic flow at RHIC in non-central heavy-ion collisions suggests that the matter created is a nearly perfect fluid with very low shear viscosity \cite{Arsene:2004fa,Back:2004je,Adams:2005dq,Adcox:2004mh}. Shear viscosity to entropy density ratio ($\eta/s$) as a measure of the fluidity is used as one of the important observables to understand the QCD medium. $\eta/s$ shows minimum at the critical point for various substances for example helium, nitrogen and water~\cite{bdm}.  Thus the measurement of $\eta/s$ can provide the required information to locate the critical end point/crossover region in the QCD phase diagram. In the framework of a relativistic kinetic theory, the shear viscosity over entropy density ratio, $\eta/s$ is given by \cite{DiasdeDeus:2012uc,Danielewicz:1984ww,Hirano:2005wx},

\begin{eqnarray}
 \eta/s \simeq \frac{T\lambda_{mfp}}{5},
 \label{e1}
\end{eqnarray}

where T is temperature, $\lambda_{mfp}$ is the mean free path calculated by using the following formula,
 
\begin{eqnarray}
 \lambda_{mfp} = \frac{L}{(1-e^{-\xi})}.
 \label{e2}
\end{eqnarray}

$L$ is the longitudinal extension of the string $\sim$ 1 fm. Now Eq.(\ref{e1}) becomes,

\begin{eqnarray}
 \eta/s \simeq \frac{TL}{5(1-e^{-\xi})}.
 \label{e3}
\end{eqnarray}

\begin{figure}
\includegraphics[height=20em]{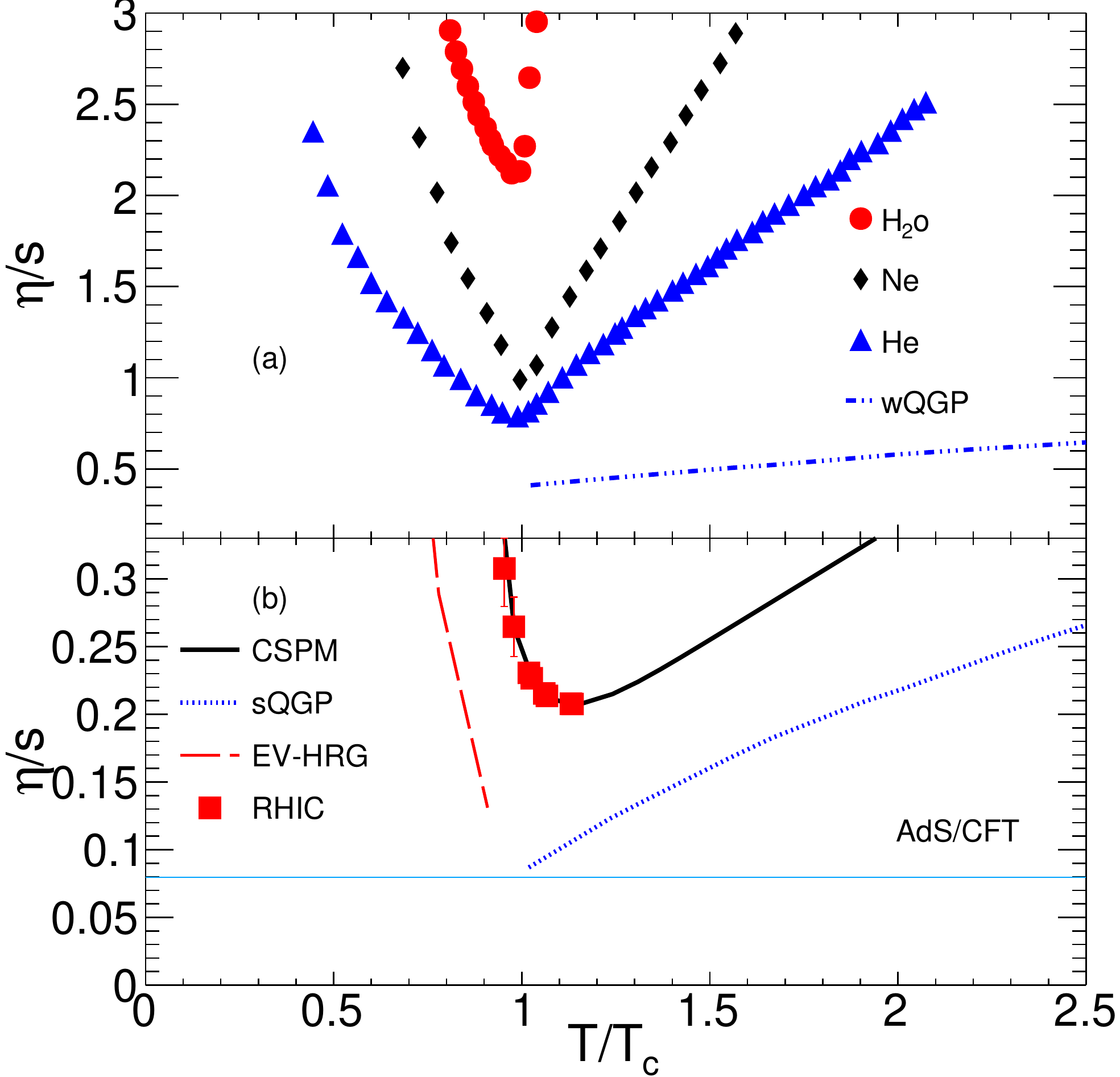}
\caption[]{(Color online) $\eta/s$ as a function of $T/T_{c}$. The red square shows the results from various RHIC energies from the CSPM. The black solid line shows the extrapolation to higher temperatures from CSPM. The solid horizontal line around  $1/4\pi$ represents the AdS/CFT limit~\cite{Kovtun:2004de}. Blue dash-dotted and blue dotted line show the results from wQGP and sQGP, respectively~\cite{Hirano:2005wx}. The red dashed line shows the calculations of EV-HRG~\cite{Tiwari:2011km}.}
\label{EtabyS}
\end{figure}

Figure \ref{EtabyS}(b) shows $\eta/s$ as a function of $T/T_{c}$. The CSPM results are shown along with  the results for weakly interacting QGP (wQGP)~\cite{Hirano:2005wx}, strongly interacting QGP (sQGP)~\cite{Hirano:2005wx}.

The lower bound of this ratio {\it i.~e.} $1/4\pi$ proposed by AdS/CFT calculations~\cite{Kovtun:2004de} is also shown in Fig.\ref{EtabyS}(b). It is observed that the matter produced in such collisions has the smallest $\eta/s$ value among any known fluids which supports the finding of the formation of perfect fluid at RHIC. For comparison purpose $\eta/s$ values for various atomic fluids are shown in Fig.~\ref{EtabyS}(a)~\cite{bdm}. The measured values of $\eta/s$ are tabulated in the Table~\ref{table1}.

\subsection{Trace anomaly}
\label{TA}
\begin{figure}
\includegraphics[height=20em]{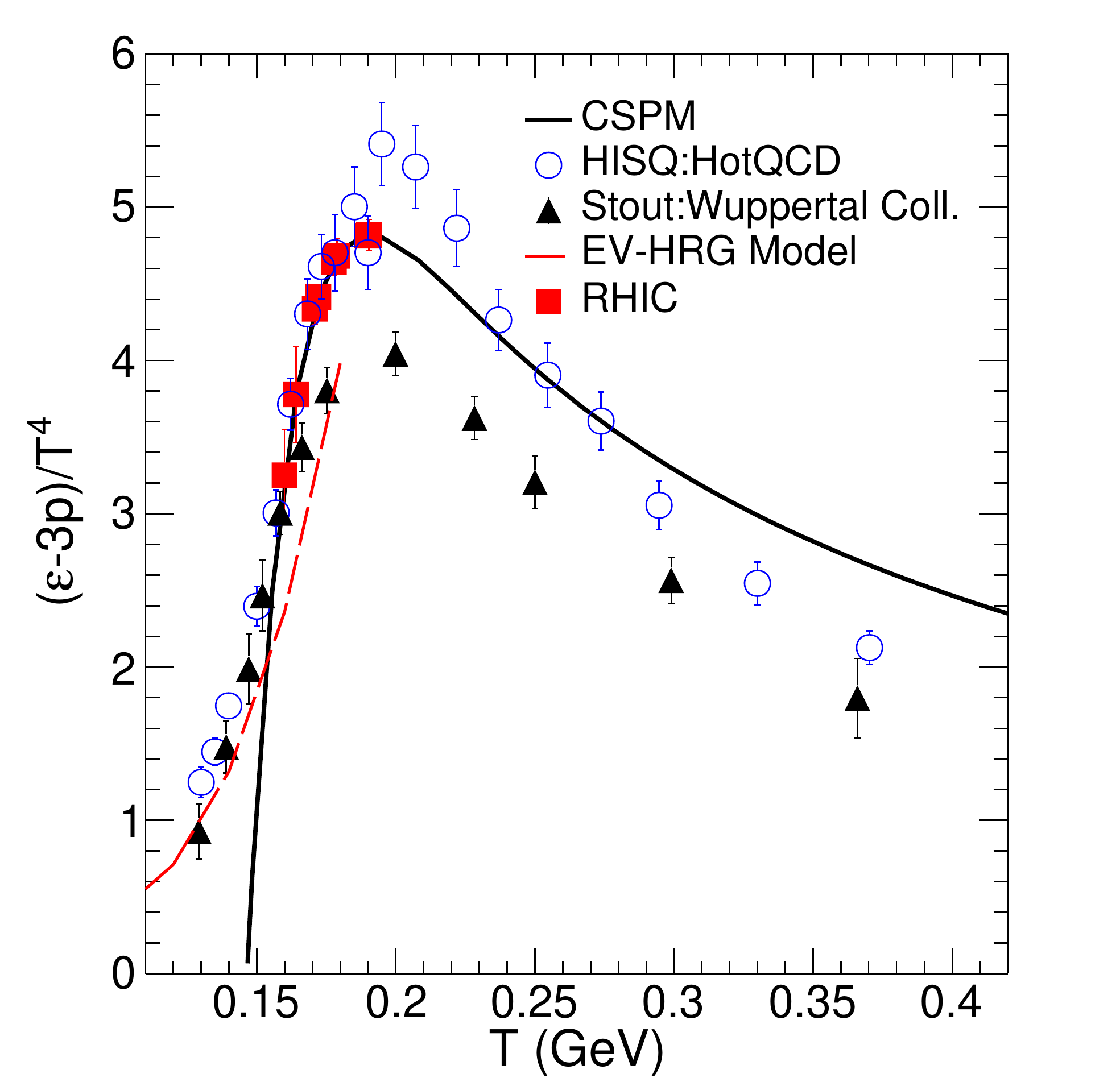}
\caption[]{(Color online) Variation of trace anomaly $\Delta = (\varepsilon-3P)/T^{4}$ with respect to temperature. Blue circles represent results from HotQCD collaboration~\cite{Bazavov:2014pvz}. Black triangle refer to Wuppertal collaboration~\cite{Borsanyi:2010cj}. The red dashed line shows the result from EV-HRG model~\cite{Tiwari:2011km}. The red squares are the results obtained from CSPM for RHIC energies and black line is the extrapolated CSPM results.}
\label{trace}
\end{figure}

Trace anomaly ($\Delta = (\varepsilon-3P)/T^{4}$) measures the deviation from the conformal behaviour, which is the trace of energy-momentum tensor, $\langle \Theta_{\mu}^{\mu}  \rangle= (\varepsilon - 3p)$. This also helps in identifying the existence of interactions in the medium~\cite{Cheng:2009zi}. The reciprocal of $\eta/s$ is in quantitative agreement with $\Delta$ for a wide range of temperatures. So, the minimum of $\eta/s$ corresponds to the maximum of $\Delta$. Figure \ref{trace} shows the variations of $\Delta$ with the temperature. We show the calculations of CSPM along with the results of lQCD from HotQCD collaboration \cite{Bazavov:2014pvz} and Wuppertal collaboration \cite{Borsanyi:2010cj}. We find that CSPM results are in close agreement with that of the HotQCD collaboration results while lie above to that of Wuppertal collaboration. The value of $\Delta$ is found to be maximum at top RHIC energy where $\eta/s$ shows minimum in CSPM calculations\cite{Phyreport}.

\subsection{Speed of Sound}
\begin{figure}
\includegraphics[height=20em]{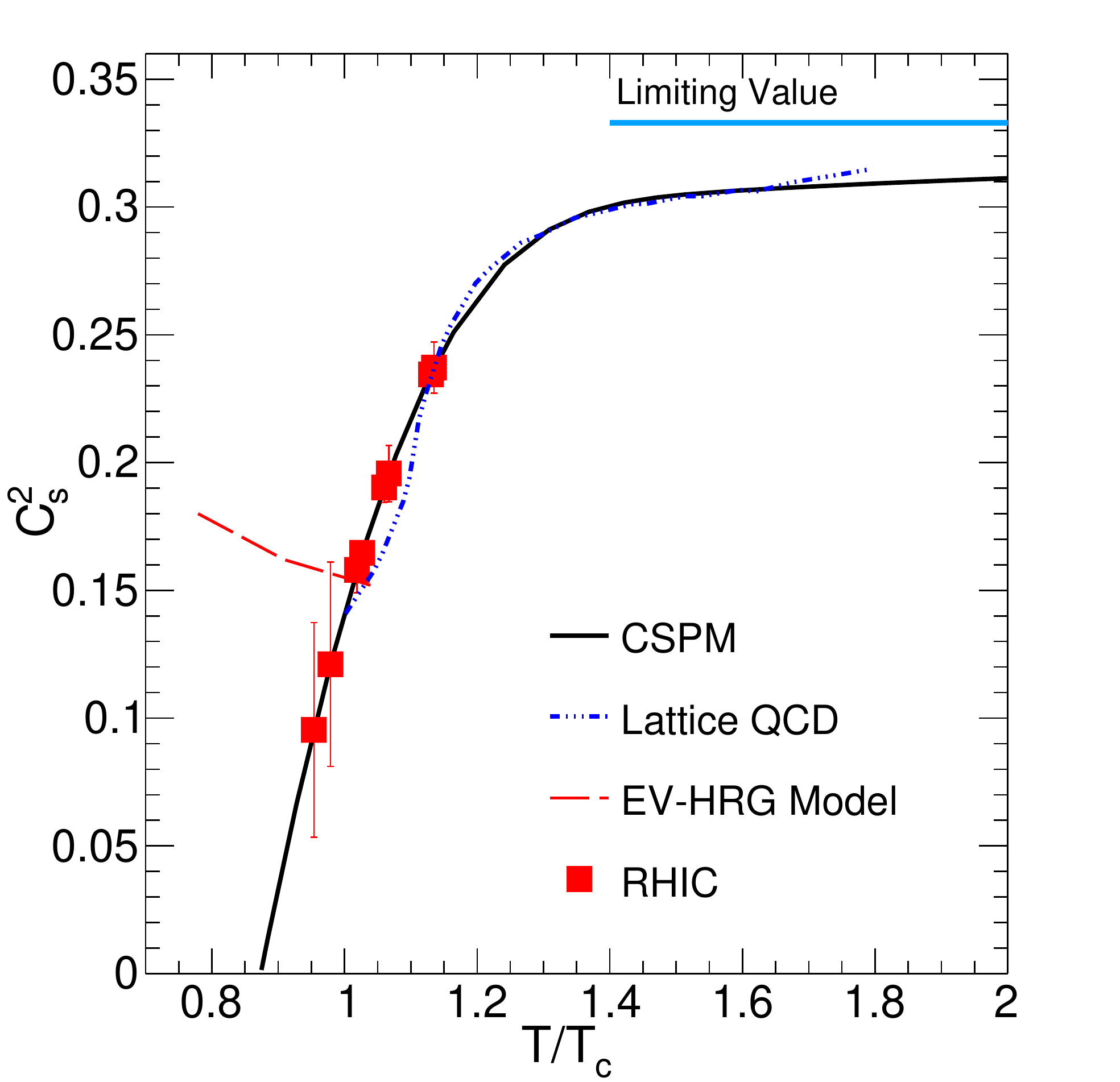} 
\caption[]{(Color online) The squared speed of sound $(C_{s}^{2})$ as a function of $T/T_{c}$. The red squares are the CSPM results at RHIC energies. The black line is the extrapolated CSPM result. Blue dash-dotted line represents  lQCD results~\cite{Bazavov:2009zn}. The red dashed line shows the result from EV-HRG~\cite{Tiwari:2011km}. }
\label{vsound}
\end{figure}

 The speed of sound is an important quantity which is related with the small perturbations produced in the medium formed in heavy-ion collisions. It explains how the change in the energy density profile of the created medium is converted into pressure gradients. In hydrodynamics, the collective expansion is observed due to pressure gradients. Using the boost-invariant Bjorken 1D hydrodynamics \cite{Bjorken:1982qr} with CSPM, the square of speed of sound ($C_{s}^2$) is calculated as \cite{Phyreport},
 
\begin{eqnarray}
&&C_{s}^2 = (-0.33) \left(\frac{\xi e^{-\xi}}{1-e^{-\xi}}-1 \right)\nonumber\\
&+&0.0191(\Delta/3)\left(\frac{\xi e^{-\xi}}{(1-e^{-\xi})^2}- \frac{1}{1-e^{-\xi}}\right),
 \label{p1}
\end{eqnarray}
where $\Delta = (\varepsilon-3P)/T^{4}$ is the trace anomaly. In Fig. \ref{vsound}, we show the variations of $C_{s}^2$ with $T/T_c$. The solid line represents the result obtained in CSPM while the blue dash-dotted line is the lattice QCD results at zero chemical potential~\cite{Bazavov:2009zn}. There is a very good agreement between these two, particularly at higher $T/T_c$ where the deviation is observed in an earlier work \cite{epjc71}. At $T/T_{c}$ =1, CSPM and lQCD results agree with the EV-HRG results shown by the red dashed line. The red squares are the results at RHIC energies. CSPM results always lie below the limiting value of $C_{s}^2$ for ideal gas which is $1/3$ at all the temperatures. These findings are expected in the case of interacting matter and suggest that the causality is respected.

\subsection{Entropy Density}
We estimate the entropy density (s) using CSPM coupled to the hydrodynamics. In CSPM, the strings interact strongly to form clusters and produce pressure and energy density at the early stages of the collisions. The expression for entropy density is,

\begin{eqnarray}
s = (1+ C_{s}^2)\frac{\varepsilon}{T}.
 \label{p1}
\end{eqnarray}

\begin{figure}
\includegraphics[height=20em]{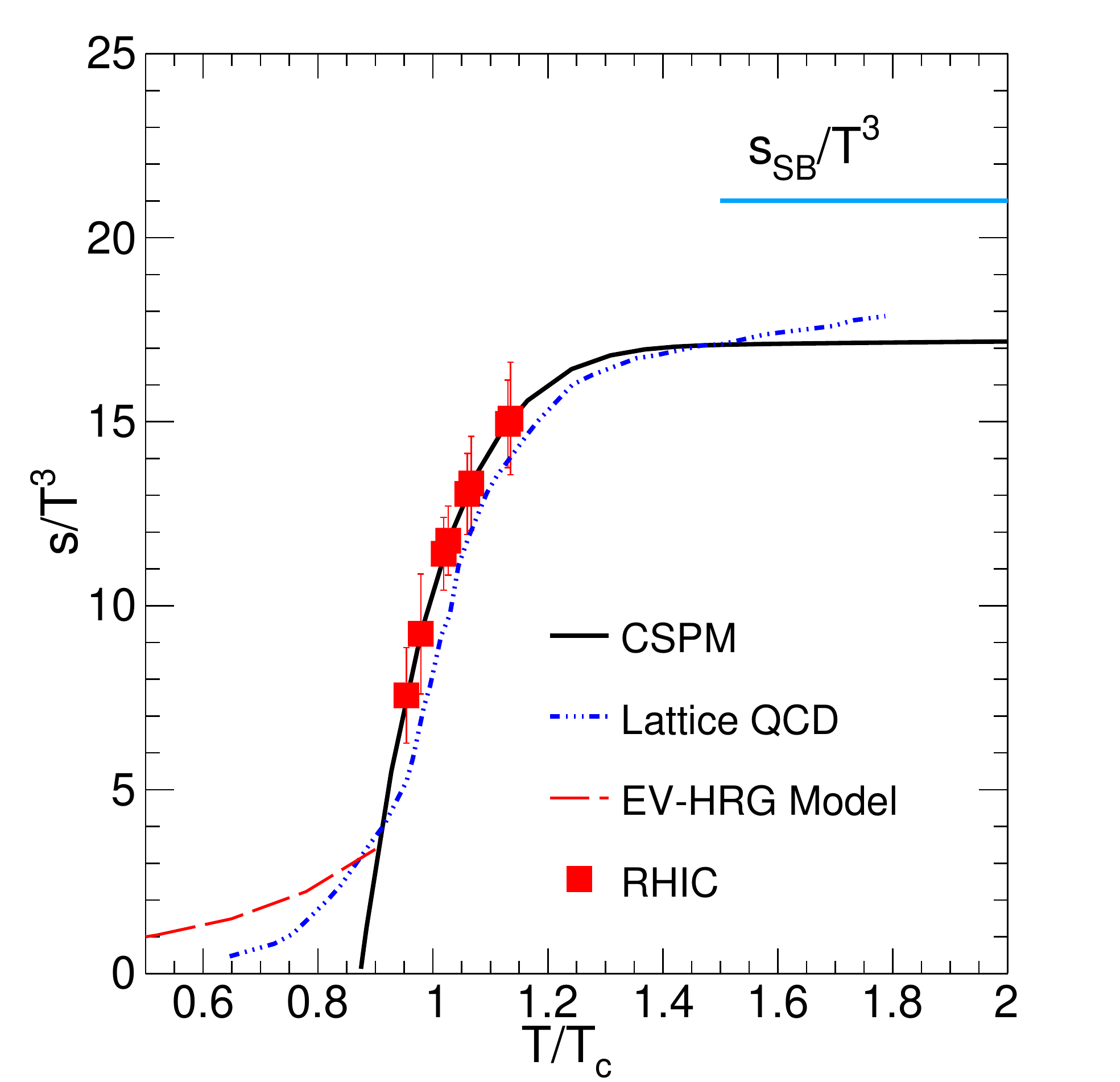}
\caption[]{(Color online) Entropy density $(s/T^{3})$ as a function of $T/T_{c}$. The red squares are the results from CSPM for RHIC energies. Blue dash-dotted line represents  the  lQCD results~\cite{Bazavov:2009zn}.  The red dashed line shows the result calculated using EV-HRG model~\cite{Tiwari:2011km}.}
\label{entropy}
\end{figure}

Figure \ref{entropy} shows the variations of scaled entropy density ($s/T^3$) with $T/T_{c}$. The solid line is the CSPM results using our parameterisation. The red squares show the results for RHIC energies. The red dashed line shows the result from EV-HRG model for a hadron gas. The solid blue line is the $s/T^3$ in the Stefan-Boltzmann limit. Again, we find a good agreement between CSPM results and lQCD data~\cite{Bazavov:2009zn} shown by the blue dash-dotted line. We find that $s/T^3$ changes rapidly with temperature. The rise of entropy density close to the phase transition temperature region, determined by the fluctuations of the chiral order parameter, is a consequence of the production of many new quark degrees of freedom \cite{Schmidt:2017bjt}.  

\subsection{Bulk Viscosity}
In the perfect fluid limit the energy density decreases with proper time due to longitudinal expansion. However, the viscosity opposes the system to perform the useful work while expanding longitudinally. In order to quantify the location of the critical point, it is very important to study the bulk viscosity to entropy density ratio ($\zeta/s$) which changes rapidly at this point. The values of $\zeta/s$ is calculated through the relation of shear viscosity and speed of sound given as follows \cite{Dusling:2011fd},

\begin{eqnarray}
 \frac{\zeta}{s} = 15\frac{\eta}{s}(\frac{1}{3} - C_{s}^2)^{2}.
 \label{p1}
\end{eqnarray}

\begin{figure}
\includegraphics[height=20em]{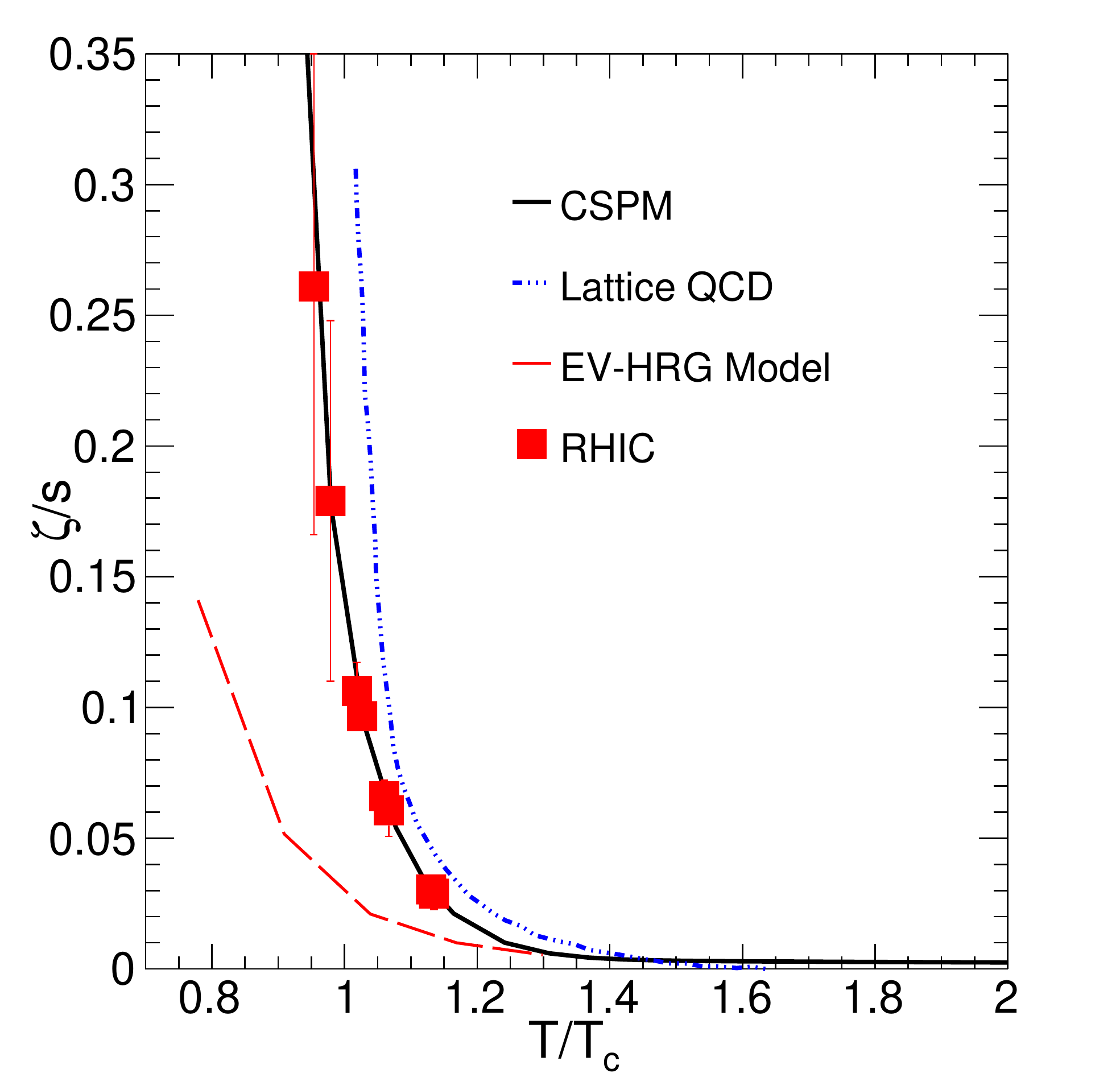}
\caption[]{(Color online) The ratio of bulk viscosity and entropy density $(\zeta/s)$ as a function of $T/T_{c}$ is shown by red squares from CSPM and the black solid line shows the extrapolated CSPM results. The blue dash-dotted line corresponds to results from lQCD \cite{Karsch:2007jc}. The red dashed line is the result of EV-HRG model~\cite{Tiwari:2011km}.}
\label{bulk}
\end{figure}

In Fig.\ref{bulk}, we show the variation of $\zeta/s$ with the temperature. We compare CSPM results with the lQCD calculations \cite{Karsch:2007jc} and again find a good agreement between these two. We observe that $\zeta/s$ is small compared to $\eta/s$  for $T/T_{c} > 1$ .

\section{Summary and Outlook}
\label{summary}
We have presented the study of equilibrium thermodynamical and transport properties such as energy density, shear viscosity, trace anomaly, speed of sound, entropy density, and bulk viscosity of the QCD matter created at RHIC energies by using the clustering of color sources phenomenology. The color suppression factor $F(\xi)$ is obtained from the  transverse momentum spectra of charged particles for most central heavy-ion collisions in order to obtain various observables. $F(\xi)$ is responsible for reduction in multiplicity and enhancement of transverse momentum. 

It is observed that the results are in excellent agreement with the lattice QCD data. The initial temperatures at RHIC energies are presented with different model predictions of chemical freeze-out temperature ($T_{ch}$), and baryon chemical potential ($\mu_B$). It is found that the initial temperatures at lower energies are very close to the chemical freeze-out temperatures. As the collision energy increases, the differences between the initial temperatures and the freeze-out temperatures increase.
The ratio of shear viscosity over entropy density has been evaluated as a function of temperature at different center-of-mass energies. $\eta/s$ decreases as a  function of energy as well as temperature upto 200 GeV and then start to increase. The whole picture is consistent with the formation of a fluid with a low $\eta/s$ ratio. We found that $\zeta/s$  diverges near critical point. The trace anomaly as a function of temperature shows similar trend as the results obtained in HotQCD collaboration. The CSPM based analysis of RHIC data from STAR show that the transition from de-confined to confined phase most likely will occur between $\sqrt{s_{NN}}$ = 11.5 and 19.6 GeV of collision energy. The analysis of the  $pp$ collisions at $\sqrt{s}$ = 7 and 14 TeV at the LHC can map events with higher temperatures and energy densities.

\section*{Acknowledgement} 
PS, SKT, SD and RNS acknowledge the financial supports from ALICE Project No. SR/MF/PS-01/2014-IITI(G) of Department of Science \& Technology, Government of India.

\end{document}